\documentclass{aa}
\usepackage{txfonts}
\usepackage[dvips]{graphicx}
\usepackage[dvips]{geometry}
\usepackage{natbib}
\bibpunct{(}{)}{;}{a}{,}

\begin{document}

\title{General Relativistic Radiative Transfer}
\author{Sebastian Knop \inst{1}\inst{2} \and Peter H. Hauschildt
  \inst{1} \and E. Baron \inst{2} } 
\institute{Hamburger Sternwarte, Gojenbergsweg 112, 21029 Hamburg, Germany \and
University of Oklahoma, 440 West Brooks, Rm 100, Norman, OK 73019-2061, USA
}
\date{Received date \ Accepted date}

\abstract
{}
{We present a general method to calculate radiative transfer including
scattering in the continuum as well as in lines in spherically symmetric systems
that are influenced by the effects of general relativity (GR). We utilize
a comoving wavelength ansatz that allows to resolve spectral lines throughout
the atmosphere.}
{The used numerical solution is an operator splitting (OS) technique that uses
a characteristic formal solution. The bending of photon
paths and the wavelength shifts due to the effects of GR are fully  taken into
account, as is the treatment of image generation in a curved
spacetime.}
{We describe the algorithm we use and demonstrate the
effects of GR 
on the radiative transport of a two level atom line in a neutron star like
atmosphere for various combinations of continuous and line scattering
coefficients.  In addition, we present grey continuum models and discuss
the effects of different scattering albedos on the emergent spectra and the
determination of effective temperatures  and radii of neutron star atmospheres.
}
{}


\maketitle

\section{Introduction}

In the last few decades much effort has been put into the modeling of
radiative transfer in relativistically moving atmospheres such as
novae and supernovae. One of the state of the art techniques to solve
this problem is the operator splitting (OS) method. This method has
been successfully used to solve radiative transfer problems with
scattering and complete treatment of non-local thermodynamic
equilibrium (NLTE) effects.

However, these sophisticated methods for treating radiative transfer
have not been used in a general relativistic environment, although the
form of the GR equation of transfer does, basically, not differ from
the special relativistic version and the OS method can be applied to
such problems.

There have been several successful attempts to model the emergent
spectra of general relativistic systems such as neutron stars.  These
models also account for magnetic fields or different surface
temperatures. However, the radiative transfer in these cases generally
solves classical plane parallel problems. See
\citet{2002nsps.conf..263Z} for a review.

Here we present an OS method of solving GR radiative transfer problems
in spherical (Schwarzschild) geometry.  Other authors have solved the
radiative transfer problem in GR.  For instance
\citet{1989ApJ...346..350S} solve the moment equations of radiative
transfer with a variable Eddington factor method. The advantage of OS
is not only that it does not depend on closure conditions but can also
solve magneto-optical transfer, what could prove to be important as
far as neutron stars are concerned.

\citet{1996ApJ...466..871Z} developed a characteristics method to solve
general relativistic radiative transport problems. They utilize the
constants of motion for the description of photon orbits that arise
due to the Killing vectors of the spherically symmetric spacetime and
use the analytical connection of an affine parameter with the radial
coordinate as well as choosing the momentum variables along the characteristics
to be constant to formulate the radiative transport equation.
Although this simplifies the equation in the case of flows such as
accretion onto black holes and neutron stars, the lack of a comoving
wavelength description forces the use of a large number of
characteristics to resolve the (in this case) angular dependent
absorption coefficients and a huge number of wavelength points to
resolve the shift of spectral lines through the atmosphere. The latter
is necessary even in the case of static general relativistic
atmospheres, e.g., in neutron stars.  

We avoid these problems by using
a comoving wavelength coordinate which explicitly accounts for the
coupling of different wavelengths. This ansatz is the important core
part of this work. Since in order to calculate detailed spectra you need
to resolve spectral lines throughout the atmosphere and you cannot afford
to use a wavelength description that depends on the layer of the atmosphere
that you are at. Since in order to perform NLTE calculations you would have to
add a significant number of wavelengths to your computation to resolve the
spectral line at hand in every layer with the desired quality.

In the following we present calculations of general relativistic
radiative line and continuum transfer with a complete treatment of
scattering.  In order to demonstrate the functionality of the method,
we chose a simple test case similar to a neutron star atmosphere.

\section{Radiative Transfer}

Lindquist found the equation for radiative transfer for a comoving metric \citep{lindquist66}.
He used the photon distribution function as variable to describe the radiation field. The works 
of \citet{1989ApJ...346..350S} and \citet{1996ApJ...466..871Z} follow this ansatz.

However, we want to utilize a description of the radiation via the specific intensity that
is suitable for our method of solution.

Rather than using the general metric used by Lindquist,
we neglect the effects of the atmosphere on the metric and use 
the Schwarzschild solution:
\begin{equation}
g_{\alpha \beta} = \left( 
\begin{array}{c c c c}
1-\frac{2 M G}{c^2 r} & 0 & 0 & 0 \\
0 & -\frac{1}{1- \frac{2 M G}{c^2 r}} & 0 & 0 \\
0 & 0 &- r^2 & 0 \\
0 & 0 & 0 & -r^2 \sin^2{\Theta}
\end{array}
\right).
\label{schwarzmetric}
\end{equation}
Since the atmospheres that are influenced by GR effects
typically have a small mass compared to the parent object, this
simplification is well justified.  
Furthermore, it is possible to calculate
the metric coefficients for spherical symmetry by integrating the
Tolman-Oppenheimer-Volkov equations and thus avoiding this approximation if
desired.

The equation of radiative transfer in the Schwarzschild metric is then found in its characteristic form:
\begin{equation} 
\frac{\partial I_\lambda}{\partial s} + a_\lambda \frac{\partial \lambda I_\lambda}{\partial \lambda} + 4 a_\lambda I_\lambda = \eta_\lambda - \chi_\lambda I_\lambda
\label{EQRTschwarz}
\end{equation}
with
\begin{eqnarray}
\frac{\partial}{\partial s} & = & \frac{\partial r}{\partial s} \frac{\partial}{\partial r} + \frac{\partial \mu}{\partial s} \frac{\partial}{\partial \mu} \\
\frac{\partial r}{\partial s} & = & \sqrt{1-\frac{2 M G}{c^2 r}} \mu\\
\frac{\partial \mu}{\partial s}  & = & \frac{1-\mu^2}{r}  \left( 1 - \frac{M G}{c^2 r - 2 M G} \right) \sqrt{1-\frac{2 M G}{c^2 r}} \\
a_\lambda & = & \sqrt{\frac{r}{r - \frac{2 M G}{c^2}}} \frac{M G}{c^2 r^2} \mu
\end{eqnarray}
This development is equivalent to the work of Lindquist and delivers no new physical 
insight besides being in the same form as the spherically symmetric special relativistic
equation of transfer \citep{mihalas80}. Hence modern operator splitting techniques to
solve radiative transfer are applicable to the problem.

However,
there is a fundamental difference from the special relativistic equation
of transfer.  The coefficient $a_\lambda$ does not change sign in
monotonic flows that describe, e.g., supernova and nova
atmospheres. In the GR case, this coefficient is linear in $\mu$ and
hence the sign of $a_\lambda$ will be different for ingoing ($\mu<0$)
and outgoing ($\mu>0$) photons.  $a_\lambda$ couples different
wavelengths and determines how the spectral features shift due to the
flow or the gravitational field. In the case of a supernova atmosphere
the different parts of the atmosphere all move away from each other so
that the direction of the wavelength shift along a ray is always the
same, hence $a_\lambda$ has the same sign along the ray.  In a
gravitational field, ingoing photons will be blueshifted and outgoing
photons will be redshifted and hence, the sign of $a_\lambda$
changes.  This presents a difficulty as the direction of flow of
information in the wavelength space is reversed along a ray and the
transfer equation is no longer an initial value problem in wavelength
space, but a 2-point boundary value problem. 

Mihalas 
already realized this in \citet{mihalas80} and 
outlined a simple ray by ray formal solution 
to this problem. However, this solution is of little use for the construction
of an approximate $\Lambda$-operator for an ALI-iteration. 
To solve this problem we use the OS method
described in detail in \citet{petereddienms3}. 

The treatment of arbitrarily flows of information in the wavelengthspace means
that every spatial point at all wavelengths can influence the intensity
at a given spatial point at a given wavelength. This forces you to use a 
matrix notation for the formal solution where you have to implement proper
boundary conditions gouverned by $a_\lambda$ at every spatial point for all
wavelengths to ensure a locally stable upwind scheme for the wavelength discretization.

The GR case is a rather simple application to this method since the flow of information
in the wavelengthspace just changes once on a ray, namely at the point of tangency or in
the case of core intersecting rays at the innermost layer.

To make this method work with the GR case you have to calculate $a_\lambda$ for every
wavelength on all points of a given characteristic and determine the appropriate discretization
of the derivative at this point. Furthermore you have to know the pathlength between two
neighboring points on a ray to be able to calculate the change of optical
depth along the ray for all wavelengths. In addition to that you must determine the angles
of intersection with a given layer for all characteristics in order to perform angular integrations.

A photon in a gravitational field not only experiences a
wavelength shift but also is deflected since it moves on a
null-geodesic for the given spacetime. Since we are employing a
characteristics based solution we can fully account for this effect.
The spatial derivatives in equation (\ref{EQRTschwarz}) describe
the geometry of our characteristics.  They are still geodesics and can
be described by an affine parameter.

In order to obtain the rays, the $\frac{\partial}{\partial s}$ part of equation
(\ref{EQRTschwarz}) has to be integrated.  For the Schwarzschild metric it is
possible to analytically describe the photon orbits in terms of constants of
the motion. In addition, we have to relate the affine parameter to the distance
along the characteristics. In the one dimensional case, the integration is
simple and fast.  Since the spacetime just outside the atmosphere will, in
general, not be flat, we extend our calculation of the characteristics into a
regime of spacetime that can be considered flat -- in our test calculations
discussed below, we chose a boundary of ten Schwarzschild-radii -- to make sure
that we calculate the spectrum from a correct set of angles that represent the
imaging of the source in curved spacetime. 

We assume vacuum conditions outside
the atmosphere and, therefore, the intensities will not change along the
characteristic outside the atmosphere, of course except for being redshifted
due to the gravitational field what is trivial to account for.

\section{The Testing Setup}

We solve the test radiative transfer problems in a spherical model
configuration with 50 radial points (layers). We assume an exponential density
structure 
\begin{equation}
\varrho(r) = \varrho_0 \exp {\frac{r - r_{\rm{out}}}{r_{\rm{scale}}}}
\label{rholaw}
\end{equation}
within the atmosphere and that the gas consists only of a simple 
two-level-atom with a wavelength independent background continuum.

For a given optical depth ($\tau$) grid we integrate the radial grid via
\begin{equation}
\label{drdtau}
\frac{\mathrm{d} r}{\mathrm{d} \tau} = - \frac{1}{\chi_\kappa}
\end{equation}
where $\chi_\kappa = \chi_0 \varrho(r)$ with $\varrho$ given by (\ref{rholaw}).
It should be noted that $\chi_\kappa$ represents only the continuum extinction
coefficient. The resulting structure is not intended to be an accurate model of
a neutron star atmosphere. However, it has the correct spatial dimensions and
we thus use it to  make predictions how GR will affect radiative transfer in a
realistic neutron star atmosphere.

\begin{figure}
\begin{center}
\includegraphics[width=\hsize]{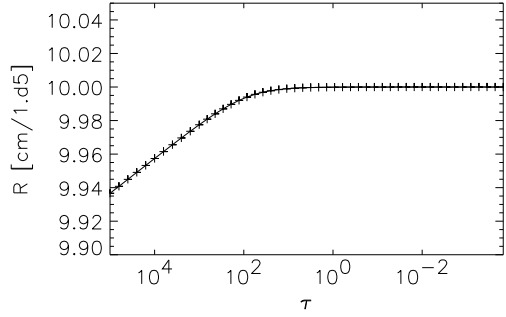}
\caption{Radius is plotted over optical depth. The optical depth was
  calculated from the wavelength independent 
continuous opacity. 
The atmosphere is about 60 meters thick, but the layers with an
optical depth around one lie just 
centimeters below the outermost layer.}
\label{fig1}
\end{center}
\end{figure}
In Fig. \ref{fig1} we plot the radial structure of the test atmosphere versus
the optical depth grid that we used for the calculations in section
\ref{secresults}.  The method of \citet{petereddienms3} makes it necessary to
provide a spatial boundary condition for the characteristics which is done via
the diffusion approximation. We generate a simple grey temperature structure
with the Hopf-function \citep{chandraradtransfer}.  To describe the single
spectral line of the two-level-atom, we define a wavelength
$\lambda_{\mathrm{line}}$ as the center of the line and use a Gaussian
profile centered on this wavelength:
\begin{equation}
\Phi(\lambda) = \frac{\omega_{\mathrm{line}}}{\sqrt{\pi}} \exp{(-\frac{\lambda - \lambda_{\mathrm{line}}}{\omega_{\mathrm{line}}})^2},
\end{equation}
with $\omega_{\mathrm{line}}$ being the width of the Gaussian.  
We describe the opacity associated with the line via:
\begin{equation}
\chi_{\mathrm{line}}(\tau,\lambda) = \chi_\kappa(\tau) R_{\mathrm{line}} \frac{\Phi(\lambda)}{\int \Phi(\lambda) \mathrm{d} \lambda},
\end{equation}
whereby the $R_{\mathrm{line}}$ factor determines the strength of the line
relative to the continuum.

It should be noted that the Gaussian width of the line is only
$0.01$ \,{\AA}. This is very small and does not represent a line width
one 
would expect in neutron star atmosphere due to the large temperatures and 
pressures present in such an atmosphere . The small width was chosen
to highlight the effects of GR on 
radiative transfer. Since the atmosphere of a neutron star is only a
few centimeters 
thick, the intrinsic wavelength shifts within the atmosphere are very small and
the GR effects can be tested best with very narrow lines. 

A detailed treatment of radiative transfer especially in the general
relativistic environment of a neutron star atmosphere is desirable, since
constraints on the mass-radius relation are needed for the understanding of
neutron star interiors and its equation of state. The constraint should be as
strict as possible and, therefore, the radiative transfer should be as
sophisticated as possible. 

Realistic models will need a multidimensional
description and have characteristics in the atmosphere that extend over larger
portions of spacetime and hence have a larger intrinsic wavelength shift.
Furthermore there may be configurations of blended lines that create a rapidly
changing opacity as seen, e.g., in the UV spectra of classical novae
\citep{novaphys}. 

Furthermore, in realistic models of accretion columns on neutron stars the
atmosphere extends over larger regions of spacetime and the intrinsic line
shifts will be much larger.  Therefore, there are physical systems where you
expect, that detailed general relativistic radiative transfer is important.
Besides this method is not limited to static atmospheres but can also be
applied to gamma-ray-bursts, accretion scenarios, or neutrino transport in
early phases of core collapse supernovae.

Since we treat the radiative transfer problem with scattering, we need
to specify 
the quantities $\epsilon_\kappa$ and $\epsilon_{\mathrm{line}}$ to
define the scattering albedo. 
The true absorption and the scattering part of the continuum opacity can now 
be expressed through:
\begin{eqnarray}
\kappa_\kappa(\tau) & = & \epsilon_\kappa \chi_\kappa(\tau) \\
\sigma_\kappa(\tau) & = & (1 - \epsilon_\kappa) \chi_\kappa(\tau) \quad,
\end{eqnarray}
whereas for the line opacity we have:
\begin{eqnarray}
\kappa_{\mathrm{line}}(\tau,\lambda) & = & \epsilon_{\mathrm{line}}\chi_{\mathrm{line}}(\tau,\lambda)\\
\sigma_{\mathrm{line}}(\tau,\lambda) & = & (1 - \epsilon_{\mathrm{line}}) \chi_{\mathrm{line}}(\tau,\lambda) \quad.
\end{eqnarray}
The total opacity can now be given as:
\begin{eqnarray}
\chi_{\mathrm{total}}(\tau,\lambda) & = & \kappa_\kappa(\tau) + \sigma_\kappa(\tau) \nonumber \\
& & \quad + \kappa_{\mathrm{line}}(\tau,\lambda) + \sigma_{\mathrm{line}}(\tau,\lambda) 
\end{eqnarray}
while the emissivity is:
\begin{eqnarray}
\eta_{\mathrm{total}}(\tau,\lambda) & = & (\kappa_\kappa(\tau)+ \kappa_{\mathrm{line}}(\tau,\lambda)  ) \; B(T(\tau))  \nonumber \\
 &  & +(\sigma_\kappa(\tau) + \sigma_{\mathrm{line}}(\tau,\lambda)  ) \;  J(\tau,\lambda). \nonumber \\
\end{eqnarray}
Note that the continuous opacity is constant over the wavelength range
of interest and that the scattering was assumed to be coherent 
for simplicity.

\section{Results}
\label{secresults}

\begin{figure}
\begin{center}
\includegraphics[width=\hsize]{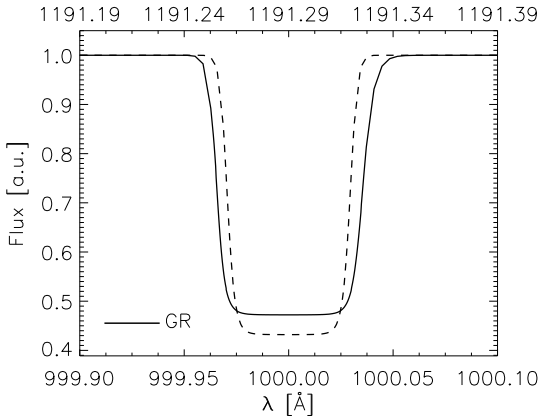}
\caption{Results for non-scattering model line and continuum.}
\label{fig2}
\end{center}
\end{figure}

\begin{figure}
\begin{center}
\includegraphics[width=\hsize]{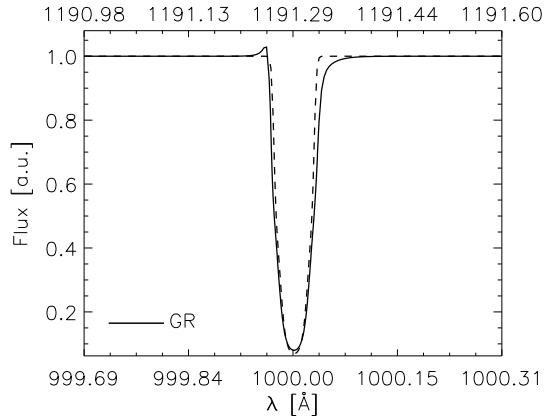}\\
\caption{The model line and the continuum are scattering with
$\epsilon_\mathrm{line} = 0.01$ and $\epsilon_\kappa = 0.01$ respectively. }
\label{fig3}
\end{center}
\end{figure}

\begin{figure}
\begin{center}
\includegraphics[width=\hsize]{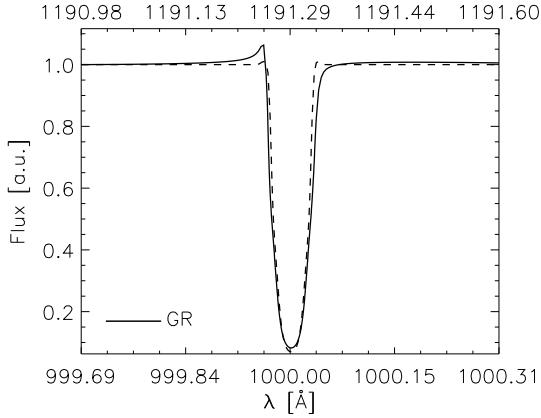}\\
\caption{The model line scatters with $\epsilon_\mathrm{line} = 0.01$ while the continuum 
scattering factor is $\epsilon_\kappa= 1.0 \times 10^{-6}$.}
\label{fig4}
\end{center}
\end{figure}

\begin{figure}
\begin{center}
\includegraphics[width=\hsize]{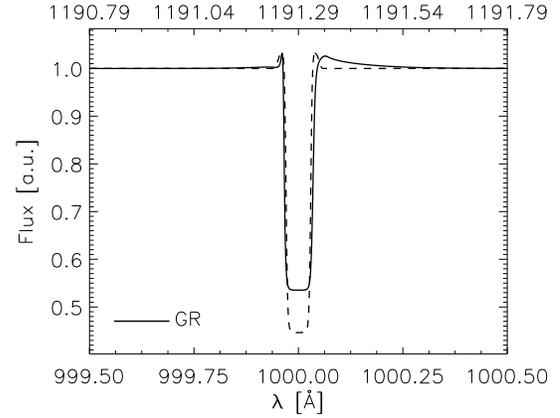}\\
\caption{The model line is nonscattering  with a completely scattering -- $\epsilon_\kappa = 0 $ -- continuum.  }
\label{fig6}
\end{center}
\end{figure}

\begin{figure}
\begin{center}
\includegraphics[width=\hsize]{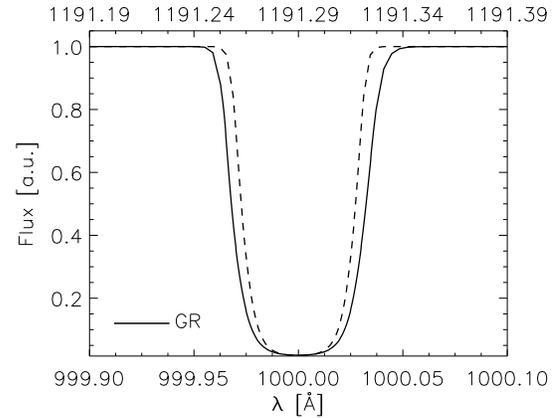}\\
\caption{The model line is completely scattering -- $\epsilon_\mathrm{line} = 0 $ -- while the continuum is
not scattering at all.}
\label{fig5}
\end{center}
\end{figure}

We calculate the emerging spectra of the model line for various
combinations of scattering parameters for the continuum and for the line 
 -- hence providing a NLTE treatment of the line transfer.
In Figs. \ref{fig2} to \ref{fig5} we always show 
two cases with different gravitational masses -- one with $\mathrm{M}
= 0$ and the other with  
$\mathrm{M} = \mathrm{M}_{\sun}$. The wavelength scale at the bottom corresponds
to the massless case and the upper scale to the one solar mass
case. The emerging 
line profiles are plotted over each other in order to be easily compared.
All calculated spectra include the full treatment of scattering unless it is
indicated that a scattering parameter was set to one.

We include the massless case in order to verify the code by comparing
the results 
to the thoroughly tested special relativistic code and 
obtained
the same results with both methods.
In addition, all other tests such as omitting the line and 
recovering a flat continuum 
for constant thermal sources, sudden changes
in the wavelength resolution, and wavelength dependent boundary
conditions produced correct results.

In Fig. \ref{fig2} the continuum as well as the line are purely
thermal. In the massless case this results in a symmetric absorption
line while in the general relativistic case for one solar mass the
line profile deforms and becomes asymmetric with an extended wing to
the red, although the effect is very small.  The line profiles are
normalized to the continuum and the equivalent widths of the two lines
are different. Although the radial structure and the run of the
opacities is exactly the same in both cases, the effective opacity is
different.  due to the factor $4 a_\lambda$ in equation
\ref{EQRTschwarz}.

In Fig. \ref{fig3} the line and the continuum are scattering with
$\epsilon_\mathrm{line} = 0.01$ and $\epsilon_\kappa = 0.01$.
In the massless case this results in a symmetric absorption profile
with line wings  
slightly in emission, while in the general relativistic case the
line profile is very asymmetric with an emission feature on the blue
side and an 
extended red absorption wing.

The model parameters used to generate Fig. \ref{fig4} resemble those
of Fig.
\ref{fig3}. The only difference is a stronger scattering in the continuum
$\epsilon_\kappa= 1.0 \times 10^{-6}$.  The strength and shape of the blue
emission depends on the scattering in the continuum. For a strong scattering
continuum even in the massless case an emission feature appears in both wings
of the line. It can be explained with the Schuster-mechanism \citep{mihalas2}.
In the relativistic case this mechanism appears to be amplified and strongly
distorted in the red wing.  To rule out any effect of line scattering, we have
calculated a model line with pure scattering ($\epsilon_\kappa=0$) in the
continuum and no scattering in the line. The result is  shown in Fig.
\ref{fig6}. The massless case shows the expected absorption profile
with the wings
in emission. In the relativistic case, the shape of the blue emission feature
does not change compared to the cases with less scattering. However, its slope
extends way farther into the blue, while on the red side of the line a very
broad emission feature extends much farther to the red, although the
peak is  not as
high as the blue peak.  In addition, we have calculated a purely scattering
($\epsilon_\mathrm{line} = 0$) model line 
combined with a purely absorptive continuum. This decouples the model line 
from the thermal pool.  The
results are shown in Fig. 
\ref{fig5}.  Due to the scattering, the line is very strong, but the
deformation of the lines wings is essentially the same as in Fig. \ref{fig2},
where the line did not scatter. 

The effects of GR for the given atmosphere are most obvious for a
scattering continuum. The resulting lineshape is strongly
asymmetric. 
Scattering in the line 
has no discernible effect on the shape of the line. The actual deformation due
to GR effects is also very small for pure linescattering.

The emergent lineshapes are related in principle to P-Cygni profiles, since
there is also a constant wavelength shift -- the dopplershift --  throughout the 
atmosphere. However there
is a fundamental difference between the two cases, since in an expanding atmosphere
where P-Cygni profiles are observed there is a point of last contact of a photon with
the atmosphere, hence the information of the dopplershift at the point
of emission in respect to the observer is conserved. In the gravitational field the
wavelength of the photon gets shifted even without any interaction with the atmosphere.

\section{Grey continuum models}

As another application to general relativistic transfer we calculated grey
continuum models. To do so we changed our wavelength resolution and omitted the
model line from our calculations ($R_{\mathrm{line}}=0$).  In the following we
present the emerging continuous spectra for varying values of
$\epsilon_\kappa$. 

The temperature structure of the model is grey with an
effective temperature of $10^4\,$K (the absolute value of the effective
temperature is not important for the testing).  The emergent spectra are nearly
blackbody spectra with only a slightly distorted overall shape as they are a
little bit broader than the blackbody.  

\begin{figure}
\begin{center}
\includegraphics[width=\hsize]{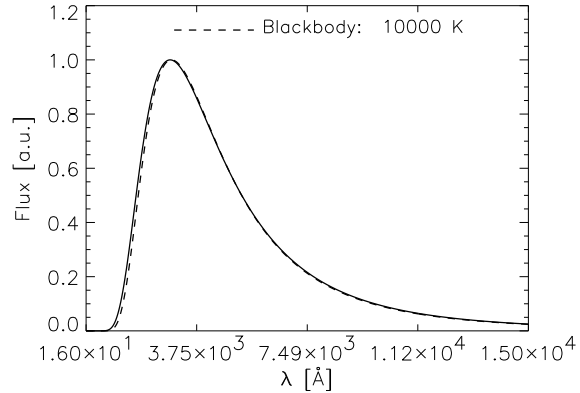}\\
\caption{The spectrum was corrected for the gravitational redshift and fitted
with a blackbody with the known effective temperature of the model -
$T_{\mathrm{eff}}= 10^4\,$K. The continuum has $\epsilon_\kappa = 1$ The
blackbody fits the model 
reasonably well.}
\label{fig9a}
\end{center}
\end{figure}

\begin{figure}
\begin{center}
\includegraphics[width=\hsize]{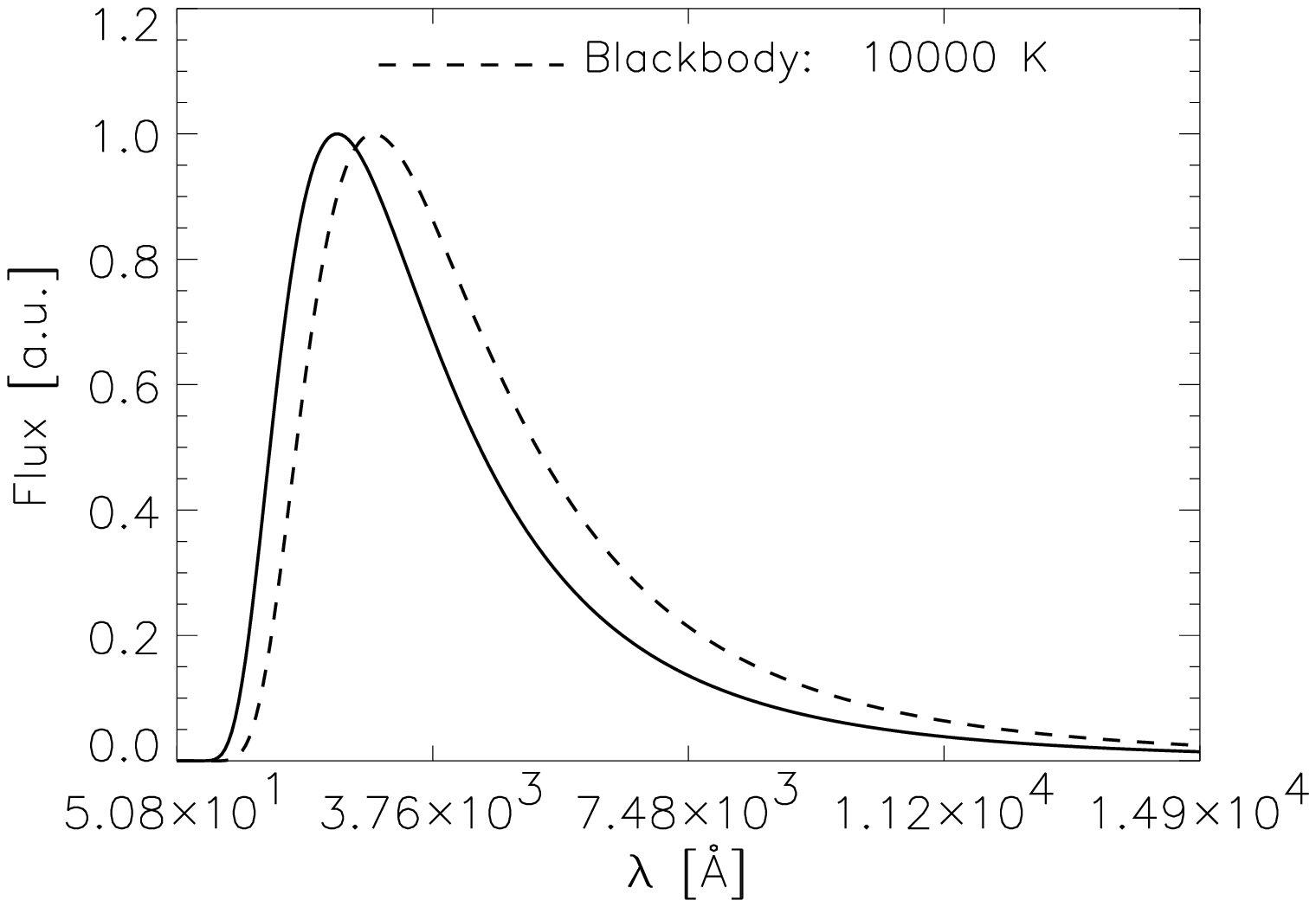}\\
\caption{The spectrum was corrected for the gravitational redshift and fitted
with a blackbody with the known effective temperature of the model -
$T_{\mathrm{eff}}= 10^4\,$K. The continuum has $\epsilon_\kappa = 0.1$ The
blackbody doesn't fit the model spectrum. The apparent temperature is higher
than the model temperature.}
\label{fig9}
\end{center}
\end{figure}

\begin{figure}
\begin{center}
\includegraphics[width=\hsize]{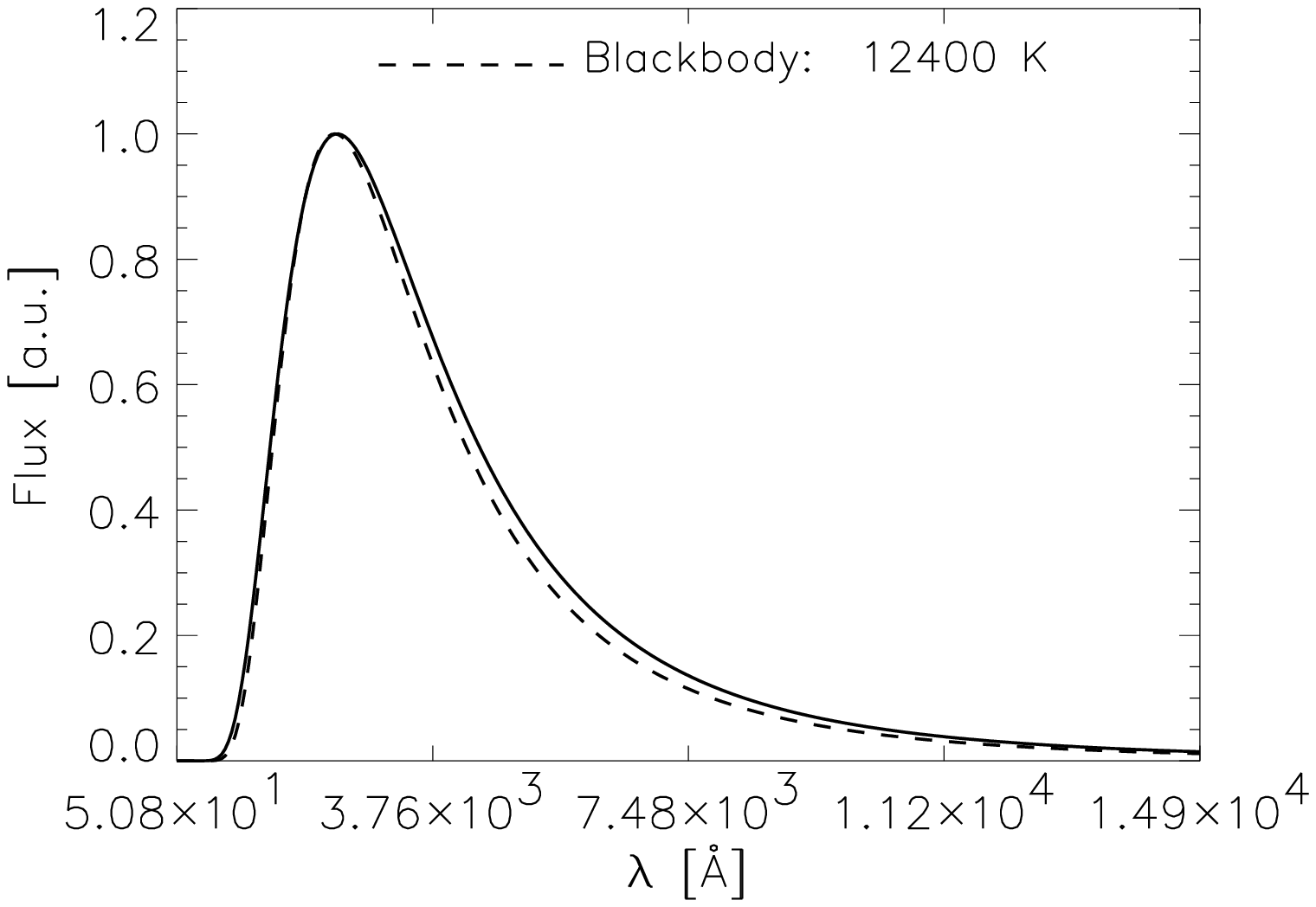}\\
\caption{The model spectrum is the same as in Figure \ref{fig9}, but this time
the temperature of the blackbody was chosen to fit the spectrum.}
\label{fig10}
\end{center}
\end{figure}

\begin{figure}
\begin{center}
\includegraphics[width=\hsize]{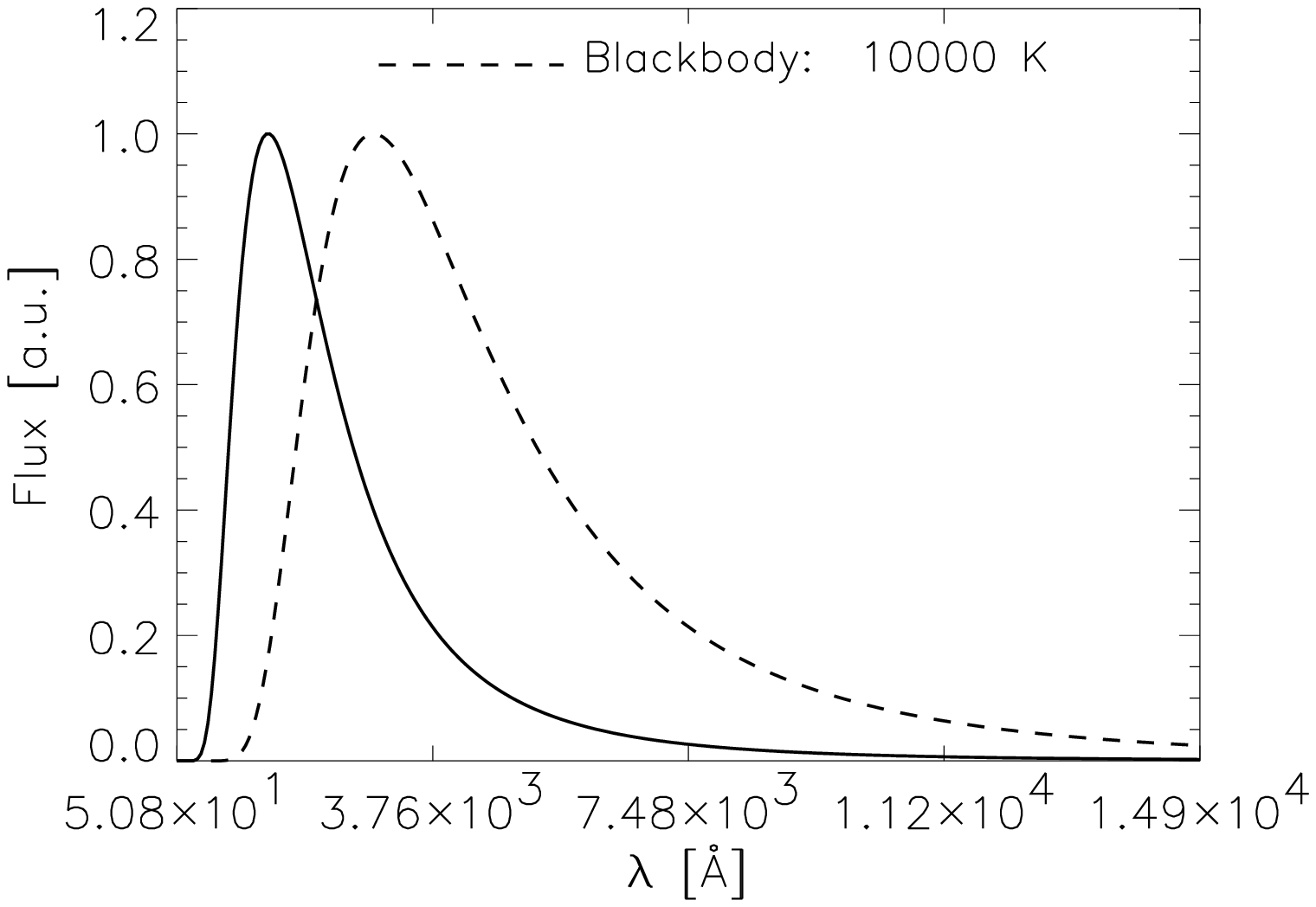}\\
\caption{The spectrum was corrected for the gravitational redshift and fitted
with a blackbody with the known effective temperature of the model -
$T_{\mathrm{eff}}= 10^4 $K. The continuum has $\epsilon_\kappa = 0.001$ The
blackbody doesn't fit the model spectrum. The apparent temperature is higher
than the model temperature.}
\label{fig11}
\end{center}
\end{figure}

\begin{figure}
\begin{center}
\includegraphics[width=\hsize]{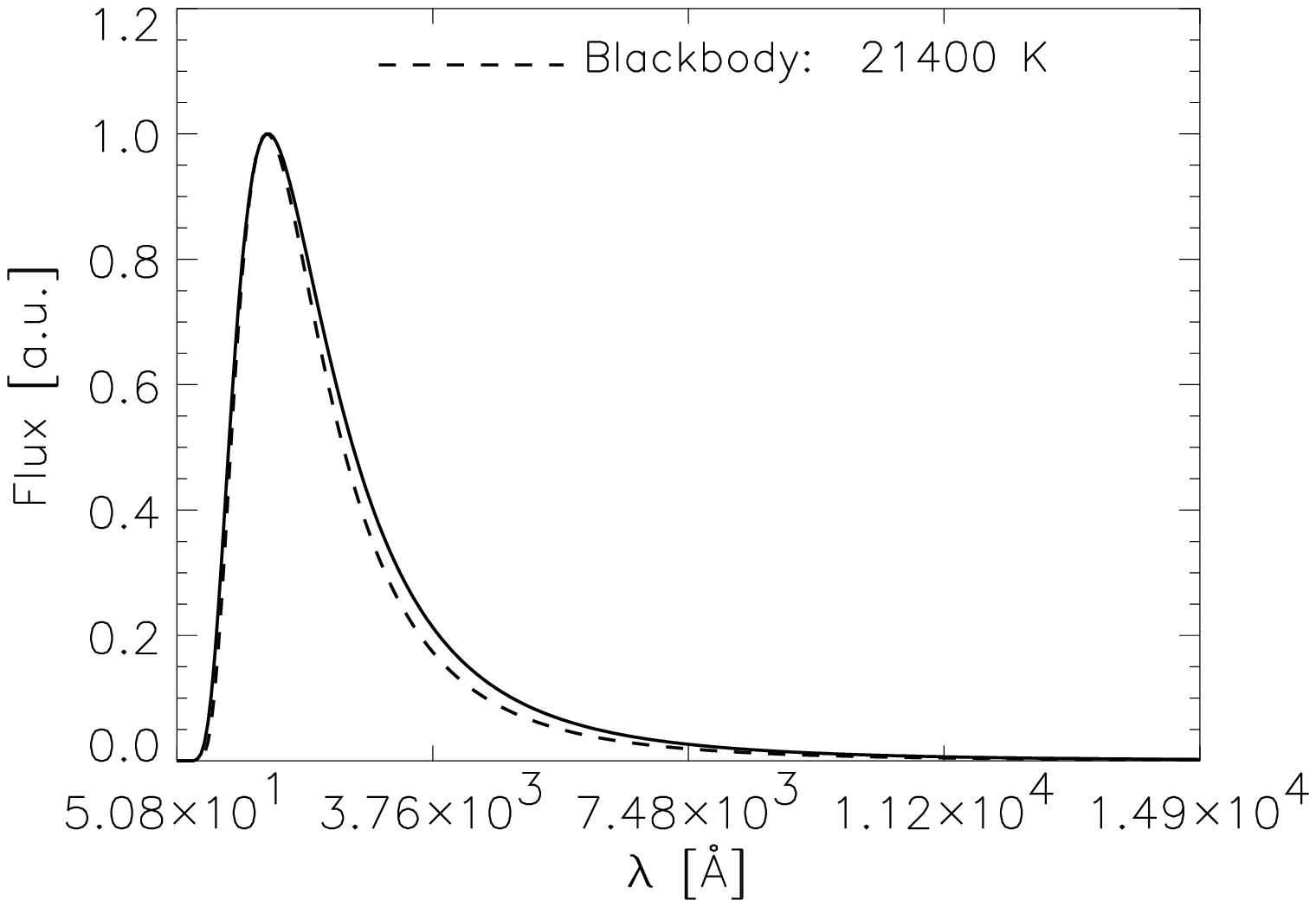}\\
\caption{The model spectrum is the same as in Figure \ref{fig11}, but this time
the temperature of the blackbody was chosen to fit the spectrum.}
\label{fig12}
\end{center}
\end{figure}

To determine the effective temperature
from an observation one would have to correct for the gravitational redshift
and then try to fit a blackbody to the spectrum. This procedure will only
deliver physically relevant results if the scattering conditions in the
atmosphere are known.  As long as all photons are absorbed and none scattered
-- $\epsilon_\kappa = 1$ -- (see Fig. \ref{fig9a}) the emergent spectrum
(corrected for gravitational redshift) is fitted with a blackbody with a
temperature that is the same as the effective temperature of the model. For
models with non-zero scattering -- $\epsilon_\kappa = 0.1$ in Fig. \ref{fig9}
and $\epsilon_\kappa = 0.001$ in Fig. \ref{fig11} -- the emerging spectra are
much bluer than the blackbody fit with the model effective temperature.  In
Fig. \ref{fig10} and Fig. \ref{fig12} we corrected the temperatures of the
blackbody fits to match the emerging spectra. The apparent temperatures are
much higher than the effective temperature of the model.  

\begin{figure}
\begin{center}
\includegraphics[width=\hsize]{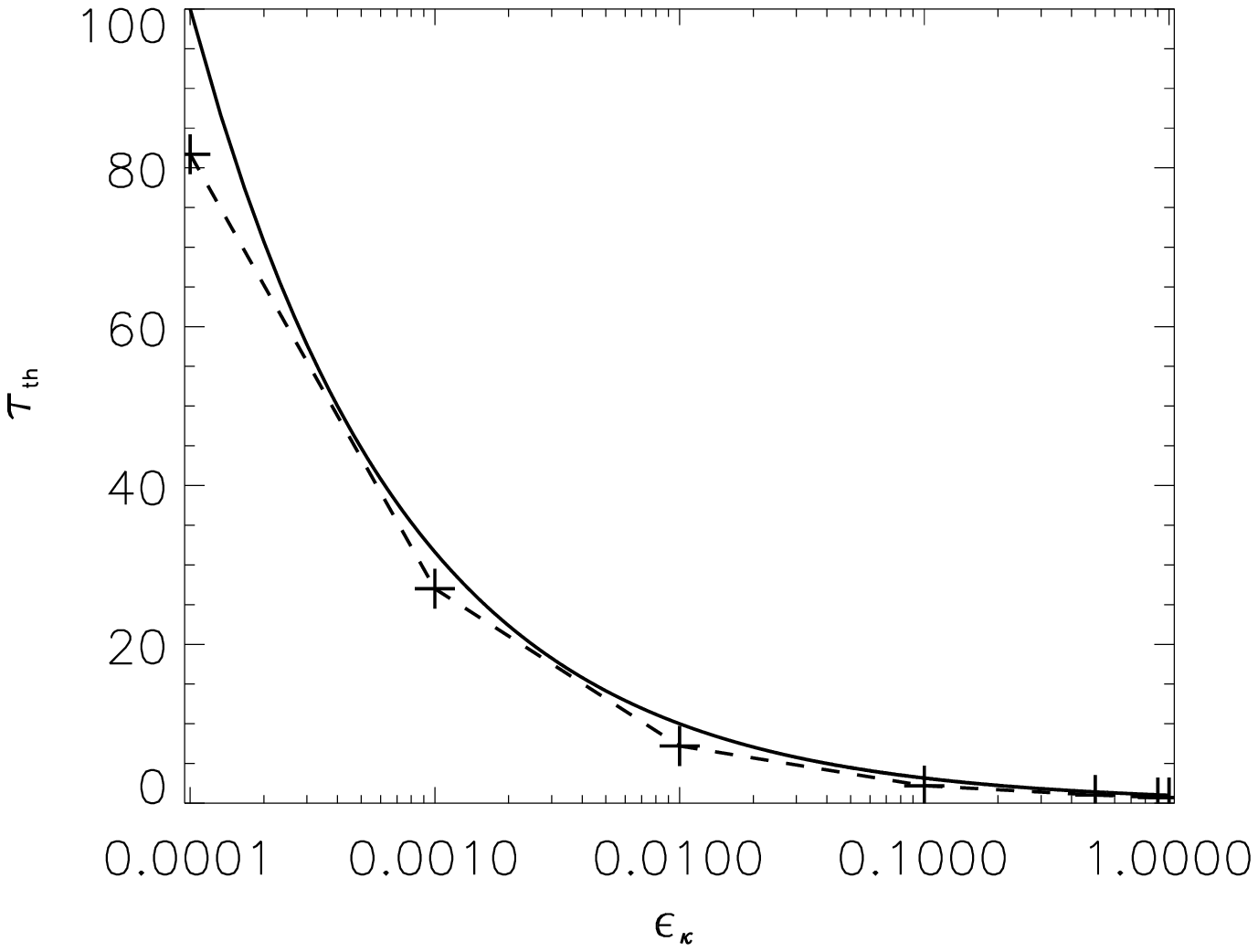}\\
\caption{Thermalisation depth is plotted over optical depth.
Since the radial structure is only known on a discrete grid the value
of the thermalisation depth for a given temperature was determined via linear interpolation.}
\label{fig8}
\end{center}
\end{figure}

To prove the
consistency of our 
results,  we plot the thermalisation depth $\tau_{\mathrm{th}}$ - the optical
depth where the $J_\lambda = B_\lambda$ - over the scattering parameter
$\epsilon_\kappa$. Since the emergent spectrum is Planckian,
$\tau_{\mathrm{th}}$ is the optical depth where the temperature of the
atmosphere equals the temperature of the blackbody fit. The results are plotted
in Fig. \ref{fig8}.


The results are consistent with the simple model $\tau_{\mathrm{th}}
=\frac{1}{\sqrt{\epsilon_\kappa}}$.  Hence, in the GR case the determination of
effective temperatures via blackbody fits or simple radiative transfer models
that neglect scattering will be inconsistent and result in systematic errors of
the effective temperatures. 
Therefore, it is desirable to solve the scattering
problem in a GR environment selfconsistently in a full blown atmospheric code with the method of solution that we
present in this paper.

\section{Conclusion}

We have developed a method to solve continuum and line radiative
transfer problems in spherically symmetric spacetimes that fully
accounts for general relativistic effects and can account
for scattering in the continuum and the lines. It uses a comoving
frame wavelength formalism that allows resolution of spectral lines
throughout the atmosphere without significantly increasing the number
of wavelength 
points.  The method was developed and tested in static neutron
star-like atmospheres, but is generally applicable to general
relativistic 
systems. Only the photon orbits and the wavelength coupling term
$a_\lambda$ would be different in other GR systems.  The test models
provide an illustration of possible results for realistic model
atmospheres for neutron stars.  The results show that the emergent
line profiles in general relativistic atmospheres cannot be described
in detail with non-relativistic radiative transfer.  The influence of
the continuous scattering opacity on the shape of the lines is large.

The apparent effective temperature of continuous spectra also depends strongly
on the strength of the scattering. Therefore, it is necessary to include the
treatment of scattering in the radiative transfer solution in order to obtain a
consistent physical model of a neutron star atmosphere and similar cases.  The
method that we have presented here is a first step to develop a thorough
treatment of general relativistic atmosphere models in 3D. It can be directly
applied to multi-level NLTE calculations of relativistic neutron star
atmospheres, which we will present in a subsequent paper.

\bibliographystyle{aa}

\end{document}